\documentclass[epjST]{svjour}

\usepackage{braket}
\usepackage[table]{xcolor}
\usepackage{dsfont}
\usepackage{booktabs}
\usepackage{geometry}
\usepackage{cite}
\usepackage{amsmath}
\usepackage{amsfonts}
\usepackage{amssymb}
\usepackage{dcolumn}
\usepackage[english]{babel}
\usepackage{epsfig}
\usepackage{graphics}
\usepackage{xcolor}
\usepackage{float}
\usepackage{multirow}
\usepackage{graphics}
\begin{document}
\title{Projectile motion of surface gravity water wave packets:\\ An analogy to quantum mechanics}
\author{Georgi Gary Rozenman\inst{1,2}\fnmsep\thanks{\email{georgiro@mail.tau.ac.il}}\and Matthias Zimmermann\inst{3,4} \and  Maxim A. Efremov\inst{3,4} \and Wolfgang P. Schleich\inst{3,4,5} \and{William B. Case}\inst{6} \and Daniel M. Greenberger\inst{7} \and Lev Shemer\inst{8} \and Ady Arie\inst{2}}

\institute{Raymond and Beverly Sackler School of Physics $\&$ Astronomy, Faculty of Exact 
Sciences, Tel Aviv University, Tel Aviv 69978, Israel \and 
School of Electrical Engineering, Iby and Aladar Fleischman Faculty of Engineering, 
Tel Aviv University, Tel Aviv 69978, Israel \and 
Institut f\"ur Quantenphysik and Center for Integrated Quantum Science and Technology (IQ$^{\mathrm{ST}}$), Universit\"at Ulm, 89081 Ulm, Germany \and 
Institute of Quantum Technologies, German Aerospace Center (DLR), 89081 Ulm, Germany \and
Hagler Institute for Advanced Study at Texas A$\&$M University, 
Texas A$\&$M AgriLife Research, Institute for Quantum Science and Engineering (IQSE), and Department of Physics and Astronomy, 
Texas A$\&$M University, College Station, TX 77843-4242, USA  \and 
Department of Physics, Grinnell College, P.O. Box 805, Grinnell, IA 50112, USA\and
City College of the City University of New York, New York City, NY 10031, USA\and
School of Mechanical Engineering, Iby and Aladar Fleischman Faculty of Engineering, Tel Aviv University, Tel Aviv 69978, Israel}

\abstract{
We study phase contributions of wave functions that occur in the evolution of Gaussian surface gravity water wave packets with nonzero initial momenta propagating in the presence and absence of an effective external linear potential. Our approach takes advantage of the fact that in contrast to matter waves, water waves allow us to measure both their amplitudes {\it and} phases.} 
\maketitle

\section{Introduction}

\label{intro}
Complex-valued probability amplitudes constitute \cite{Feynman_RMP_1948} the elementary building blocks of quantum mechanics \cite{Feynman_QMPI_1965,Landau} and are at the core of every interference phenomenon distinguishing the microscopic from the macroscopic world. Conventional wisdom states that it is impossible to observe the phase of a \textit{single} probability amplitude. In contrast, we report measurements of such phases in an analogue system of quantum mechanics. 

Here we take advantage of the fact, that the time evolution of a wave function in quantum mechanics is in many aspects analogous \cite{Rozenman2019b} to that of paraxial optical beams \cite{Born}, surface gravity water wave pulses \cite{Fu2015a,Fu2015b-1,Fu2016a,Rozenman2020a,Berry_AB_Effect}, and underwater acoustic beams \cite{Segev}. Interestingly, in one of these systems, a global quantum-mechanical phase, the so-called Kennard phase \cite{Kennard,T3paper}, has recently \cite{Rozenman2019a} been measured\footnote{We emphasize that the Kennard phase has also been observed \cite{Amit_PRL_2019} in a Stern-Gerlach atom interferometer.}. 

In the present article, we extend this work and study the influence of an initial momentum on the propagation of ballistic Gaussian surface gravity water waves in the presence and absence of an effective linear potential\footnote{For studies on ballistic wave packets in systems analogous to surface gravity water wave pulses we refer to Refs.  \cite{Hayata,Christodoulides2008,Chang2001}.}. 
Our measurements focus on the phases of these wave packets induced by the initial momentum, and represent the first step towards a deeper understanding of the Galilean transformation \cite{Greenberger_PRL_2001,Greenberger_Varenna_2019} in quantum mechanics. A much more detailed publication dedicated to this topic is presently in preparation.

Our article is organized as follows. In Section \ref{sec: surface_gravity_waves} we prepare the ground for the presentation of our experimental work by recalling the analogy between the Schr\"odinger equation and
the wave equation of surface gravity water waves. Here we focus on the problem of a particle in the presence of a constant force and provide the associated expressions for the amplitude and phase of an initial Gaussian wave packet.

Section \ref{sec: experimental_results} summarizes our experimental results measuring the phase of Gaussian wave packets moving in the presence and absence of a linear potential with and
without an initial momentum. In this way we demonstrate that it is possible to  retrieve the phase of this classical analogue of the Schrödinger equation. 

We conclude in Section \ref{sec: conclusion} by summarizing our main results and providing an outlook.

\section{Analogy to surface gravity water waves}
\label{sec: surface_gravity_waves}
In the present section we first briefly summarize the analogy between surface gravity water waves and quantum mechanics. We then present the expressions for the amplitude and phase of Gaussian wave packets propagating in a linear potential. Here we focus especially on the role of an initial momentum.

\begin{table}
\begin{center}
\caption{Correspondence of the relevant quantities for quantum mechanical and surface gravity water waves as discussed in more detail in Ref. \cite{Rozenman2019b}.}\vspace*{0.2cm}
 \label{tab: Correspondence_Quantum_Mechanics_Hydrodynamics}
\begin{tabular}{ ll|ll}
\toprule
\multicolumn{2}{c|}{\textbf{quantum mechanical waves}} &  \multicolumn{2}{c}{\textbf{surface gravity water waves}}    \\ 
\midrule
 position & $z$ & rescaled time (comoving frame) & $\tau$ \\ 
\rowcolor{black!10} time & $t$ &rescaled propagation coordinate & $\xi$ \\ 
 wave function & $\psi(z,t)$ & amplitude envelope & $A(\tau,\xi)$ \\
 \rowcolor{black!10} mass & $m$ & - & $1/2$ \\ 
reduced Planck constant & $\hbar$ & - & 1 \\
\rowcolor{black!10} imaginary unit & $i$& -& -$i$\\ 
linear potential &$-Fz$ & effective linear potential & $-F\tau$\\
\rowcolor{black!10} initial momentum &$p_0$ & effective initial momentum & $p_0$\\
  \bottomrule
\end{tabular}
\end{center}
\end{table}

The Schr\"odinger equation of a quantum particle described by the wave function $\psi=\psi(z,t)$ in a linear potential $-Fz$ takes the form
\begin{equation}
i\hbar \frac{\partial}{\partial t}\psi(z,t)=\left(-\frac{\hbar^2}{2m}\frac{\partial^2}{\partial z^2}-Fz\right)\psi(z,t)\,.
\end{equation}

According to Table \ref{tab: Correspondence_Quantum_Mechanics_Hydrodynamics} the corresponding dynamical equation for the normalized amplitude envelope $A= A(\tau, \xi)$ of a surface gravity water wave \cite{Mei,ShemerD} reads 
\begin{equation}
 \label{NLSE3}
 i\frac{\partial }{\partial\xi}A(\tau,\xi)=\left(\frac{\partial^2}{\partial\tau^2}+F\tau\right) A(\tau,\xi)
\end{equation}
in the {\it comoving frame} with the group velocity $c_g$. 

The scaled dimensionless variables $\xi$ and $\tau$ 
are related to the propagation coordinate $x$ and the time $t$ by $\xi\equiv \varepsilon^2 k_0x$ 
and $\tau\equiv \varepsilon \omega_0\left(x/c_g-t\right)$. The carrier wave number $k_0$ and the angular carrier frequency $\omega_0$ 
satisfy the deep-water dispersion relation $\omega_0^2=k_0g$ with 
$g$ being the gravitational acceleration, and define the group velocity $c_g\equiv \omega_0/2k_0$. The parameter $\varepsilon\equiv k_0 a_0$ characterizing the wave steepness is assumed to be small in the linear regime, that is $\varepsilon \ll 1$, where $a_0$ is the maximum amplitude of the envelope.

The effective potential $-F\tau$ in Eq. (\ref{NLSE3}) is determined \cite{Mei} by the derivative $\left(\partial\Phi/\partial\tau\right)|_{Z=0}$ of the external dimensionless velocity potential $\Phi\equiv\phi/(\omega_0 a_0^2)$ at the free surface given by the dimensionless vertical coordinate $Z=0$ with
$Z\equiv\varepsilon k_0 z$. Hence, we can create the potential $F\tau\equiv 4\varepsilon\left(\partial\Phi/\partial\tau\right)|_{Z=0}$ by an externally operating water pump.

The complex amplitude envelope $A\equiv|A|\exp(i\varphi)$ determines the variation in time and space of the surface elevation
\begin{equation}
 \eta(t,x)\equiv a_0|A(t,x)|\cos[k_0 x-\omega_0 t+\varphi(t,x)]
\end{equation}
including the carrier wave.

For Gaussian wave packets with non-zero effective initial momentum $\Omega_0$ the temporal surface elevation
\begin{equation}
 \label{Soliton1}
 \eta(t,0)\equiv a_0 \exp\left(-\frac{t^{2}}{t_{0}^{2}}\right)\cos(-\omega_{0}t+\Omega_0t)
\end{equation}
is prescribed by the wave maker at $x=0$, where $t_0$ denotes the initial pulse size.

The resulting time evolution of $A$ is given by 
\begin{equation}
 \label{amplitude-G-result}
 |A(\tau,\xi)|=\left(\frac{1}{1+\xi^2/\xi_s^2}\right)^{\frac{1}{4}}
 \exp\left[-\frac{\left(\tau - 2p_0\xi  -F\xi^2\right)^2}{\tau_0^2\left(1+\xi^2/\xi_s^2\right)}\right]
\end{equation}
and
\begin{align}
 \label{phase-G-result}
 \varphi(\tau,\xi)&= \frac{1}{2}\arctan\left(\frac{\xi}{\xi_s}\right)-\frac{\xi}{\xi_s}\frac{\left(\tau-2p_0\xi -F\xi^2\right)^2}{\tau_0^2\left(1+\xi^2/\xi_s^2\right)}-F(\tau-2 p_0\xi)\xi+\frac{F^2\xi^3}{3}-p_0\tau+p_0^{2}\xi-p_0 F \xi^{2},
\end{align}
where $\tau_0\equiv\varepsilon\omega_0 t_0$ and $\xi_s\equiv \tau_0^2/4$. 

The deviation $\Omega_0$ from the carrier frequency $\omega_0$ together with the wave steepness $\varepsilon$ determine the initial dimensionless momentum $p_0$ of the wave packet by the relation
\begin{equation}
p_0 \equiv \frac{\Omega_0}{\varepsilon\omega_0}. 
\end{equation}

The first term in Eq. (\ref{phase-G-result}) corresponds to the Gouy phase \cite{GouyPhaseshift}. Moreover, we refer to the global phase cubic in $\xi$ and quadratic in $F$ expressed by the fourth term as the Kennard phase~\cite{Kennard,T3paper}. 

It is instructive to relate the amplitude envelope $A_0=A_0(\tau,\xi)$ and the phase $\varphi_0=\varphi_0(\tau,\xi)$ with a vanishing initial momentum, that is $p_0=0$, to the one given by Eqs. \eqref{amplitude-G-result} and \eqref{phase-G-result} with the same initial profile $A(\tau,0)= A_0(\tau,0) \exp\left(-i p_0\tau\right)$, but an additional phase proportional to $p_0$. Indeed, according to Eqs. \eqref{amplitude-G-result} and \eqref{phase-G-result} the amplitude $|A_0|$ experiences a shift in the argument $\tau$ reminiscent of a Galilean transformation, that is
\begin{equation}
|A(\tau,\xi)|= |A_0(\tau-2 p_0\xi,\xi)|.
\end{equation}
However, the phase
\begin{equation}
\varphi(\tau,\xi)=\varphi_{0}(\tau-2 p_0\xi,\xi)-p_0\tau+p_0^2\xi-p_0F\xi^2
\label{eq: transformation_phase}
\end{equation}
contains apart from the same shift in the argument of $\varphi_0$ additional terms, determined by $p_0$. 

\section{Experimental results}
\label{sec: experimental_results}

In order to measure the amplitude $|A|$ and the phase $\varphi$ of surface gravity water wave pulses, we have conducted a series of experiments with water waves moving in a computer-controlled time-dependent water flow. The experimental facility is a $5\,{\rm m}$ long water tank with a computer-controlled wave-maker discussed in detail in Ref. \cite{Rozenman2019a}. The velocity of the homogeneous flow increases linearly in time and is induced by a computer-controlled water pump. In this way we can generate flow velocities with a large enough value of $F$, needed to observe a parabolic trajectory in a space-time diagram.

\begin{figure}
\includegraphics[width=1.0\textwidth]{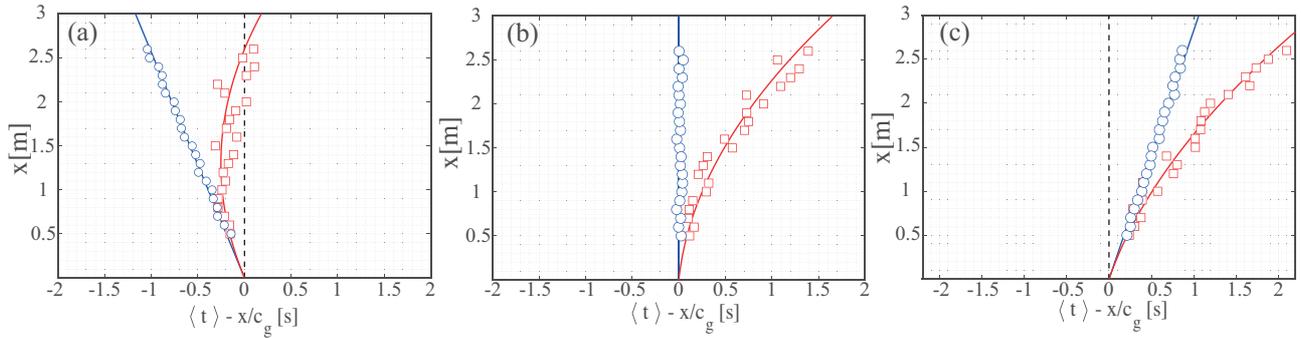}\\
  \caption{\label{Fig1} (color online) Spacetime trajectories of Gaussian wave packets propagating freely (blue lines) and in a linear potential (red lines) for the initial momenta (a) $\Omega_{0}=2\,{\rm rad/s}$, (b) $\Omega_{0}=0\,{\rm rad/s}$, and (c) $\Omega_{0}=-2\,{\rm rad/s}$. In all cases $k_0=20\,{\rm m^{-1}}$, $a_{0}=3.0\,{\rm mm}$ ($\varepsilon=0.06$), $c_{g}=0.35\,{\rm m/s}$, and $t_0=0.8\,{\rm s}$. The mean values $\langle t \rangle (x)$ (blue circles and red squares) represent the temporal coordinate, Eq. (\ref{mean value}). The corresponding solid lines are quadratic fits $\langle t\rangle (x)=a_1 x+a_2 x^2$ determining the group velocity $c_g=\left(a_1+2\Omega_0/g\right)^{-1}$ and the effective force $F\equiv -(\omega_0/\varepsilon^3k_0^2)a_2$ in Eq. (\ref{NLSE3}).}
\end{figure}

We extract the trajectory of the center-of-mass of the wave packet in the laboratory frame by calculating the temporal mean value 
\begin{equation}
 \label{mean value}
 \langle t \rangle (x)\equiv \frac{\int_{-\infty}^{+\infty}t |\eta(t,x)|^2 dt}{\int_{-\infty}^{+\infty}|\eta(t,x)|^2 dt}.
\end{equation}
at a distance $x$ from the input.

Figure \ref{Fig1}(a) shows for the positive initial momentum $\Omega_{0}= 2\,{\rm rad/s}$ the center-of-mass motion of a Gaussian wave packet that propagates freely (blue line) and moves to the left. In the presence of a linear potential (red line) the same initial wave packet propagates at the beginning of the test section in the direction opposite to the acceleration, reaches zero momentum at approximately $x=1.5\,{\rm m}$, followed by a movement to the right. 
In Fig. \ref{Fig1}(b) we depict the case of zero initial momentum, $\Omega_{0}= 0\,{\rm rad/s}$. The center-of-mass of the wave packet is stationary in the comoving frame when launched without the external potential (blue line), but accelerates and follows a parabolic trajectory when the external potential is applied (red line). For a negative momentum, $\Omega_{0}=-2\,{\rm rad/s}$ shown in Fig. \ref{Fig1}(c), the wave packet is launched to the right, accelerates in the presence of a linear potential (red line) and reaches even a larger group velocity. The black dashed lines shown for comparison are calculated for a stationary wave packet (in the comoving frame). The difference between the solid and dashed lines proves that the ballistic wave packet follows a different trajectory. 

For each set of measurements presented in Fig. \ref{Fig1}, we perform a quadratic fit of $\langle t \rangle (x)$, that is $\langle t\rangle (x)=a_1x+a_2x^2$, and obtain the coefficients $a_{1}$ and $a_{2}$. Indeed, for $\Omega_0=2\,{\rm rad/s}$, $\Omega_0=0\,{\rm rad/s}$, and $\Omega_0=-2\,{\rm rad/s}$ we have  $a_{1}=2.44\,{\rm s/m}$, $a_{1}=2.86\,{\rm s/m}$, and $a_{1}=3.23\,{\rm s/m}$, respectively. These values of $a_1$ give rise to $c_g=\left(a_1+2\Omega_0/g\right)^{-1}=0.351\,{\rm m/s}$, $c_g=0.35\,{\rm m/s}$, and $c_g=0.354\,{\rm m/s}$, and are in a good agreement with $c_{g}=0.35\,{\rm m/s}$ calculated using the water-wave dispersion relation.

In the presence of a linear potential we have obtained for the three values of $\Omega_0$ depicted in Fig. \ref{Fig1} the same value $a_2=0.15\,{\rm s/m^{2}}$ which yields $F=-(\omega_0/\varepsilon^3k_0^2)a_2=-24.4$. To validate these measurements, a Pitot tube was used to measure the velocity of the water flow $2\,{\rm cm}$ beneath the surface. For more details see Ref. \cite{Rozenman2019a}.

\begin{figure}[H]
\includegraphics[width=1.0\textwidth]{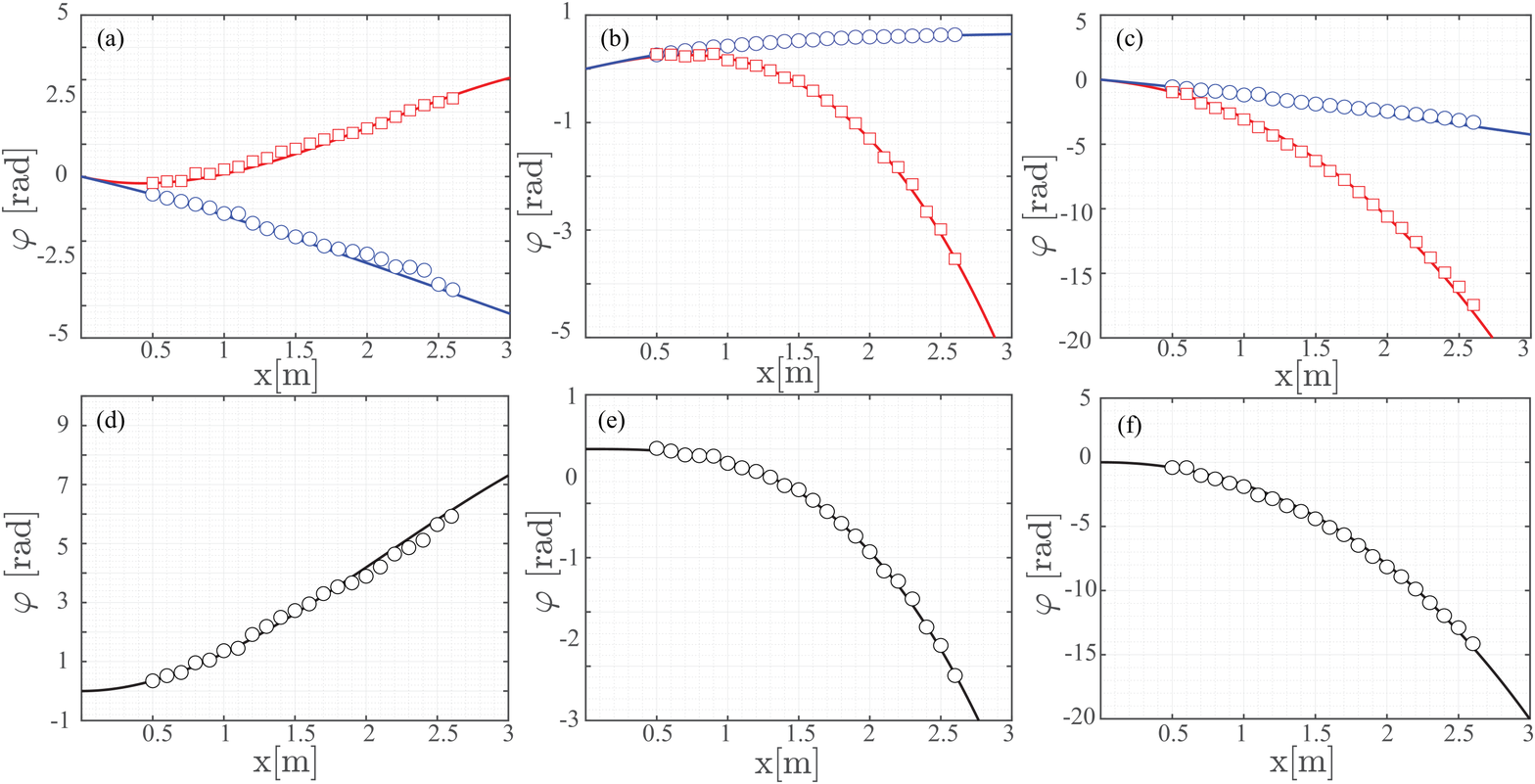}\\
  \caption{\label{Fig2} (color online) 
Phases (a)-(c) of Gaussian wave packets at the maximum without (blue circles) and with (red squares) external flow compared to the prediction given by Eq. (\ref{phase}) and represented by solid lines for the initial momenta (a) $\Omega_{0}=4\,{\rm rad/s}$, (b) $\Omega_{0}=0\,{\rm rad/s}$, and (c) $\Omega_{0}=-4\,{\rm rad/s}$. In all cases $k_0=20\,{\rm m^{-1}}$, $a_{0}=6.0\,{\rm mm}$ ($\varepsilon=0.12$), $c_{g}=0.35\,{\rm m/s}$, and $t_0=0.8\,{\rm s}$. Here we have obtained $F=-3.86$ in the presence of an external flow. The difference of the red squares and blue circles is presented as circles in (d)-(f) together with the solid black line given by $-(2/3)F^2\xi^3-2p_0 F\xi^2$ for the corresponding value of the initial momentum $\Omega_0$.}
\end{figure}

To extract the phase of the wave packet, we apply the Hilbert transform to convert the real signal determined by the surface elevation  $\eta=\eta(t,x)$ to a complex one and define the phase as the arctangent of the ratio between its imaginary and real part. For more details we refer to Refs. \cite{Rozenman2019b,Fu2015b-1,Rozenman2019a,Hilbert,HilbertMathworks}.

In this way we obtain the phase $k_0 x-\omega_0 t+\varphi(t,x)$ at the maximum 
\begin{equation}
 \label{tau_cm}
 \tau_{cm}=2p_0\xi + F\xi^2
\end{equation}
of the Gaussian amplitude $|A(t,x)|$ given by Eq. (\ref{amplitude-G-result}).

After removing the carrier phase $k_{0}x-\omega_0 t$, we present the remaining phase in Figs. \ref{Fig2} (a)-(c) by blue circles (without external flow) and red squares (with external flow), together with the blue and red solid curves corresponding to the analytical expression 
\begin{equation}
 \label{phase}
 \varphi(\tau_{cm},\xi) = \frac{1}{2}\arctan\left(\frac{\xi}{\xi_s}\right)-\frac{2}{3}F^2\xi^3-p_0^{2}\xi-2p_0 F \xi^{2}
\end{equation}
derived from Eq. (\ref{phase-G-result}) at $\tau=\tau_{cm}$, given by Eq. (\ref{tau_cm}). Here we have used the initial momenta (a) $\Omega_0=4\,{\rm rad/s}$, (b) $\Omega_0=0\,{\rm rad/s}$, and (c) $\Omega_0=-4\,{\rm rad/s}$. 

In the absence of a linear potential, that is $F=0$, the phase $\varphi(\tau_{cm},\xi)$, Eq. \eqref{phase}, only depends on the absolute value of $p_0$. For this reason, the blue curves in Figs. \ref{Fig2} (a) and \ref{Fig2} (c) are identical.

Due to measuring the phase at the maximum of the wave packet the coefficient of the Kennard phase has changed from $1/3$ to $-2/3$. Moreover, according to Eq. \eqref{eq: transformation_phase}, the last two terms of Eq. \eqref{phase} are related to the phase contributions induced by the effective initial momentum $p_0$, although their prefactors have been altered for the same reason. 

From the difference of the blue and red circles in Figs. \ref{Fig2} (a)-(c) we can extract those phase contributions in Eq. \eqref{phase} that are proportional to $F$. We display the resulting phase $-(2/3)F^2\xi^3-2p_0 F\xi^2$ by a solid black line in Figs. \ref{Fig2} (d)-(f) for the corresponding initial momenta $\Omega_0$.

\section{Conclusions}
\label{sec: conclusion}

In conclusion, we have observed the projectile motion of Gaussian surface gravity water wave packets propagating in the presence and absence of a linear potential. In addition, we have measured the phases of these wave packets at the \textit{maximum} of their amplitudes for three different initial momenta. 

However, it would be desirable to resolve the phase as a function of both variables $\xi$ \textit{and} $\tau$. Unfortunately, this task is challenging as it requires the phase extraction from a small signal with a low error. 

We emphasize that our measurements of the phases of surface gravity water waves, serving as an analogue system for quantum mechanics, are intimately connected to the Galilei transformation of the Schr\"odinger equation. Although the Schr\"odinger equation depends \cite{Greenberger_PRL_2001,Greenberger_Varenna_2019}   on the frame of reference, a unitary transformation ensures its form invariance.
At the very heart of this observation is the difference between canonical and kinetic momentum. In quantum mechanics it is the canonical momentum that manifests itself in the phase of the wave function. A measurement of the corresponding phase, such as reported in our article, gives us a tool to distinguish active from passive motion. The observation of such a difference would  have far-reaching consequences. Unfortunately, due to the limited space our article can only allude to these aspects. A more detailed discussion will appear in a future publication.

\section*{Acknowledgements}

We thank Tamir Ilan and Anatoliy Khait for technical support and assistance. This work is funded by DIP, the German-Israeli Project Cooperation (AR 924/1-1, DU 1086/2-1) supported by the DFG, and the Israel Science Foundation 
(Grant Nos. 1415/17 and 508/19). G.G.R. is grateful for the oppurtunity to participate in the FQMT'19 conference, during which he was inspired to study several new topics. M.Z. thanks the German Space Agency (DLR) with funds provided by the Federal Ministry for Economic Affairs and Energy (BMWi) due to an enactment of the German Bundestag under Grant Nos. DLR 50WM1556, 50WM1956. M.A.E. is thankful to the Center for Integrated Quantum Science and Technology (IQ$^{\mathrm{ST}}$) for its generous financial support. W.P.S. is grateful to Texas A$\&$M University for a Faculty Fellowship at the Hagler Institute for Advanced Study at the Texas A$\&$M University as well as to the Texas A$\&$M AgriLife Research. The research of the IQ$^{\mathrm{ST}}$ is financially supported by the Ministry of Science, Research and Arts Baden-W\"urttemberg.


\begin{thebibliography}{99}

\bibitem{Feynman_RMP_1948} R.P. Feynman, Rev. Mod. Phys. {\bf 20}, 367 (1948).

\bibitem{Feynman_QMPI_1965}R.P. Feynman and A.R. Hibbs, {\it Quantum Mechanics and Path Integrals} (McGraw-Hill, New York, 1965).

\bibitem{Landau} L.D. Landau and E.M. Lifshitz, {\it Quantum Mechanics: Non-Relativistic Theory} (Pergamon Press, 1977).


\bibitem{Rozenman2019b} G.G. Rozenman, S. Fu, A. Arie, and L. Shemer, MDPI-Fluids {\bf 4}(2), 96 (2019).


\bibitem{Born} M. Born and E. Wolf, {\it Principles of Optics} (Cambridge University Press, 2002).

\bibitem{Fu2015a} S. Fu, Y. Tsur, J. Zhou, L. Shemer, and A. Arie, Phys. Rev. Lett. {\bf 115}, 034501 (2015).
 
\bibitem{Fu2015b-1} S. Fu, Y. Tsur, J. Zhou, L. Shemer, and A. Arie, Phys. Rev. Lett. {\bf 115}, 254501 (2015).
 
\bibitem{Fu2016a} S. Fu, Y. Tsur, J. Zhou, L. Shemer, and A. Arie, Phys. Rev. E {\bf 93}, 013127 (2016).

\bibitem{Rozenman2020a} G.G. Rozenman, L. Shemer, and A. Arie, Phys. Rev. E {\bf 101}, 050201(R) (2020).


\bibitem{Berry_AB_Effect} M.V. Berry, R.G. Chambers, M.D. Large, C. Upstill, and J.C. Walmsley, Eur. J. Phys. {\bf 1}, 154 (1980). 
 
\bibitem{Segev} U. Bar-Ziv, A. Postan, and M. Segev, Phys. Rev. B {\bf 92}, 100301 (2015).

\bibitem{Kennard} E.H. Kennard, Z. Phys. {\bf 44}, 326 (1927); J. Frank. Inst. {\bf 207}, 47 (1929).

\bibitem{T3paper} M. Zimmermann, M.A. Efremov, A. Roura, W.P. Schleich, S.A. DeSavage, J.P. Davis, A.
Srinivasan, F.A. Narducci, S.A. Werner, and E.M. Rasel, Appl. Phys. B {\bf 123}, 102 (2017).

\bibitem{Rozenman2019a} G.G. Rozenman, M. Zimmermann, M.A. Efremov, W.P. Schleich, L. Shemer, and A. Arie, Phys. Rev. Lett. {\bf 122}, 124302 (2019).

\bibitem{Amit_PRL_2019} O. Amit, Y. Margalit, O. Dobkowski, Z. Zhou, Y. Japha, M. Zimmermann, M.A. Efremov, F.A. Narducci, E.M. Rasel, W.P. Schleich, and R. Folman, Phys. Rev. Lett. {\bf 123}, 083601 (2019).


\bibitem{Hayata} K. Hayata, Y. Tsuji, and M. Koshiba, J. Appl. Phys. {\bf 72}, 2912 (1992).

\bibitem{Christodoulides2008} G.A. Siviloglou, J. Broky, A. Dogariu, and D.N. Christodoulides, Opt. Lett. {\bf 33}(3), 207 (2008).

\bibitem{Chang2001} J. Wolf, Y. Pan, G.M. Turner, M.C. Beard, C.A. Schmuttenmaer, S. Holler, and R.K. Chang, Phys. Rev. A {\bf 64}, 023808 (2001).


\bibitem{Greenberger_PRL_2001} D.M. Greenberger, Phys. Rev. Lett. {\bf 87}, 100405 (2001).

\bibitem{Greenberger_Varenna_2019} D.M. Greenberger, in \textit{Foundations of quantum theory}, Proceedings of the International School of Physics ``Enrico Fermi'' Course 197, edited by E.M. Rasel, W.P. Schleich, and S. W\"olk (IOS Press, Amsterdam, 2019).


\bibitem{Mei} C.C. Mei, {\it The Applied Dynamics of Ocean Surface Waves} (Wiley-Interscience, 1983).


\bibitem{ShemerD} L. Shemer and B. Dorfman, Nonlinear Process. Geophys. {\bf 15}, 931 (2008).

\bibitem{GouyPhaseshift} S. Feng and H.G. Winful, Opt. Lett. {\bf 26}, 485 (2001).

\bibitem{Hilbert} F.W. King, {\it Hilbert Transforms}, Volume 1 (Cambridge University Press, 2009).
\bibitem{HilbertMathworks} 
MATLAB `Hilbert Transform' package 
(https://www.mathworks.com/help/signal/ug/hilbert-transform.html).




\end{thebibliography}
\end{document}